\documentclass[pra,amsfonts,amsmath,twocolumn,superscriptaddress,nofootinbib]{revtex4}

\usepackage{color}
\usepackage{graphicx} 
\usepackage{epstopdf}%

\usepackage{mathpazo}      

\newcommand {\be} {\begin{equation}} 
\newcommand {\ba}{\begin{eqnarray}} 
\newcommand {\ee} {\end{equation}} 
\newcommand{\ea} {\end{eqnarray}}

\renewcommand{\epsilon}{\varepsilon}

\preprint{MKPH-T-11-03}

\begin{document}

\title{Excitation of an Atom by Twisted Photons}

\author{Andrei Afanasev}

\affiliation{Department of Physics,
The George Washington University, Washington, DC 20052, USA}

\author{Carl E.\ Carlson}

\affiliation{Department of Physics, The College of William and Mary in Virginia, Williamsburg, VA 23187, USA}

\author{Asmita Mukherjee}

\affiliation{Department of Physics, Indian Institute of Technology, Powai, 
Mumbai 400076, India}

\date{\today}

\begin{abstract}
Twisted photon states, or photon states with large ($> \hbar$) angular momentum projection in the direction of motion, can photoexcite atomic final states of differing quantum numbers.  If the photon symmetry axis coincides with the center of an atom, there are known selection rules that  require exact matching between the quantum numbers of the photon and the photoexcited states.  The more general case of arbitrarily positioned beams relaxes the selection rules but produces a distribution of quantum numbers of the final atomic states that is novel and distinct from final states produced by plane-wave photons. Numerical calculations are presented using a hydrogen atom as an example.
\end{abstract}

\maketitle

\section{Introduction and Motivation}	\label{sec:intro}

The fact that circularly polarized photons carry an angular momentum {$\hbar$} was predicted theoretically and demonstrated experimentally in a seminal experiment by Beth in 1936 \cite{Beth36}. It was also realized \cite{Heitler} (Appendix) that the electromagnetic wave can carry orbital angular momentum in the direction of its propagation if it is constrained in the transverse plane, as in the waveguides. Much later, in 1992, Allen and collaborators suggested \cite{Allen:1992zz} that a special type of light beams that can propagate in vacuum, called Laguerre-Gaussian, predicted as non-plane wave solutions of Maxwell equations, can carry large angular momentum $ J_z \gg \hbar$ associated with their helical wave fronts. At a quantum level such beams can be described in terms of "twisted photons" \cite{molina2007nature}. This concept can be also extended to beams of particles, and electrons in particular \cite{Uchida10}.

Presently a lot of activity is focused around interactions of the twisted photons with macroscopic objects in optical tweezers, or with Bose-Einstein condensates, see Ref. \cite{Yao11} for a review and a comprehensive list of references on the subject. Methods to produce "twisted" light include spiral phase plates, computer-generated holograms \cite{Yao11}, via synchrotron radiation of electrons in a helical undulator \cite{Sasaki08, Afanasev11}, or in a free-electron laser \cite{Hemsing11}. Significant interest in the twisted photons is due to non-binary nature of the information that can be encoded by them compared to plane-wave photons of helicity ${\pm \hbar}$.  Recent theoretical work \cite{Jentschura:2010ap} shown that one can generate twisted photons with high energies of several GeV via Compton back-scattering of laser photons on an energetic electron beam, making such photon beams relevant for nuclear and particle physics. 

An important question is, to what extent absorption of the twisted photons by atoms or nuclei is different from the plane-wave photons? Recent work by Picon et al \cite{Picon10} demonstrated that during photoionization of atoms, the knocked-out electrons carry angular momenta that reproduce the angular momentum of the incoming photons. The reference \cite{Picon10} deals only with a special case in which the photon beam's symmetry axis coincides with a center of an atom. In another recent publication \cite{Davis13} the authors analyzed elastic scattering of the twisted photons on a hydrogen atom, again with a restriction that  the atom is located at the center of the optical vortex. We considered a more general case of arbitrary positioned beams and considered photoexcitation of bound states with different quantum numbers in a hydrogen atom. Our main interest was in the excitation of internal degrees of freedom, rather than in the linear motion of the entire atom in the field of electromagnetic wave.

After presenting theoretical formalism for excitation of an atom by twisted photons, we point out novel effects caused by their unique features. Our arguments are further corroborated by theoretical calculations showing that a significant fraction of the atomic levels excited by the twisted photons could not be otherwise produced by plane-wave photons.

\section{Basic Formulae}			\label{sec:one}


The twisted photon definition here follows Serbo and Jentschura~\cite{Jentschura:2010ap,Jentschura:2011ih}, although with a more field theory based viewpoint. Another possibility would be to quantize a Laguerre-Gaussian laser mode considered in the original work by Allen et al. \cite{Allen:1992zz} ,
but our main conclusions will not be affected by this choice.

A twisted photon moving in the $z$-direction is 
\begin{align}
\label{eq:twisteddefinition}
| \kappa m_\gamma k_z \Lambda \rangle &= \int \frac{d^2k_\perp}{(2\pi)^2} 
	a_{\kappa m_\gamma}(\vec k_\perp) | \vec k, \Lambda \rangle	\nonumber\\
&= \sqrt{\frac{\kappa}{2\pi}} \  \int \frac{d\phi_k}{2\pi} (-i)^{m_\gamma} e^{im_\gamma\phi_k}  \,
	|\vec k, \Lambda\rangle
\end{align}
where $|\vec k, \Lambda\rangle$ are plane wave states, or momentum eigenstates with fixed longitudinal component $k_z$ and fixed magnitude transverse component,
\be
a_{\kappa m_\gamma}(\vec k_\perp) = (-i)^{m_\gamma} e^{im_\gamma\phi_k} \sqrt{\frac{2\pi}{\kappa}}
	\delta(\kappa - |\vec k_\perp |)	\,.
\ee
The twisted photon state can thus be viewed as a superposition of plane wave states where the momenta form a cone in momentum space with a fixed pitch angle
\be
\theta_k = \arctan\left( \frac{ |\vec k_\perp| }{ k_z } \right)		\,,
\ee
and varying azimuthal angle weighted by a phase $e^{im_\gamma\phi_k}$.

The normalization is
\be
\langle \kappa' m_\gamma' k'_z \Lambda'   | \kappa m_\gamma k_z \Lambda \rangle
	= 2\pi \, 2\omega \delta(k_z-k'_z) \delta(\kappa-\kappa') 
	\delta_{m_\gamma m_\gamma'} \delta_{\Lambda\Lambda'}
\ee
for  $\langle \vec k' \Lambda'  | \vec k \Lambda \rangle 
= (2\pi)^3 2\omega \delta^3(\vec k - \vec k') \delta_{\Lambda\Lambda'}$, and $\omega = |\vec k|$.

Using the photon field operator $A^\mu(x)$,
the wave function of a plane wave photon is
\be
\label{eq:planewave}
\langle 0 | A^\mu(x) | \vec k, \Lambda \rangle 
	= \epsilon^\mu_{\vec k,\Lambda} e^{-ikx}
\ee
and so the wave function of the twisted photon is
\begin{align}
\label{eq:twistedwave}
\mathcal A^\mu_{\kappa m_\gamma k_z \Lambda}(x)
	&= \langle 0 | A^\mu(x) | \kappa m_\gamma k_z \Lambda \rangle	\nonumber\\
&= \sqrt{\frac{\kappa}{2\pi}} \  \int \frac{d\phi_k}{2\pi} (-i)^{m_\gamma} e^{im_\gamma\phi_k}  \,
	\epsilon^\mu_{\vec k,\Lambda} e^{-ikx}	\,.
\end{align}
In cylindrical coordinates this is
\begin{align}
\mathcal A^\mu_{\kappa m_\gamma k_z \Lambda}(x)
&= e^{-i(\omega t - k_z z)}	\nonumber\\
& \times
\sqrt{\frac{\kappa}{2\pi}} \  \int \frac{d\phi_k}{2\pi} (-i)^{m_\gamma} e^{im_\gamma\phi_k}  \,
	\epsilon^\mu_{\vec k,\Lambda} e^{i \vec k_\perp{\cdot}\vec x_\perp},
\end{align}
so that the twisted photon in coordinate space has a self-reproducing 2D wave front moving forward at a speed less that the normal speed of light.

The wave front can be given explicitly with help of the Jacobi-Anger formula
\be
e^{i \vec k_\perp{\cdot}\vec x_\perp}
	= \sum_{n= -\infty}^\infty i^n e^{in(\phi_\rho - \phi_k)}  J_n(\kappa\rho)
\ee
where $\phi_\rho$ is the azimuthal angle in coordinate space, $\rho=|\vec x_\perp|$, and $J_n$ is the Bessel function,  and with
\be
\label{eq:epsilonexpand}
\epsilon^\mu_{\vec k \Lambda} = 
	e^{-i\Lambda\phi_k} \cos^2\frac{\theta_k}{2} \eta^\mu_\Lambda
	+ e^{i\Lambda\phi_k} \sin^2\frac{\theta_k}{2} \eta^\mu_{-\Lambda}
	+ \frac{\Lambda}{\sqrt{2}} \sin\theta_k \, \eta^\mu_0
\ee
where the $\eta$'s are constant vectors,
\be
\eta^\mu_{\pm 1} = \frac{1}{\sqrt{2}}  \left( 0,\mp 1,-i,0 \right)	\,,
\quad \eta^\mu_0 =  \left( 0,0,0,1 \right)	\,;
\ee
the photon polarization vector phase is like the Trueman-Wick~\cite{Trueman:1964zzb} phase convention.  Then
\begin{align}
\label{eq:twistedwf}
\mathcal A^\mu_{\kappa m_\gamma k_z \Lambda}(x) &= e^{-i(\omega t - k_z z)}	
\sqrt{\frac{\kappa}{2\pi}}  \,	\Bigg\{
	\frac{\Lambda}{\sqrt{2}} e^{im_\gamma\phi_\rho} \sin\theta_k
	J_{m_\gamma}(\kappa\rho) \, \eta^\mu_0			\nonumber\\[1ex]
& \quad + i^{-\Lambda} e^{i(m_\gamma-\Lambda)\phi_\rho}  \cos^2\frac{\theta_k}{2} 
	J_{m_\gamma-\Lambda}(\kappa\rho) \, \eta^\mu_\Lambda	\nonumber\\[1ex]
& \quad + i^{\Lambda}  e^{i(m_\gamma+\Lambda)\phi_\rho}  \sin^2\frac{\theta_k}{2} 
	J_{m_\gamma+\Lambda}(\kappa\rho) \, \eta^\mu_{-\Lambda}
	\Bigg\}	\,.
\end{align}

As an aside, if we were to write the photon wave function for a plane-wave photon of helicity $\Lambda$, it would be like the above, possibly with some differences of normalization choice, but with pitch angle $\theta_k\to 0$ (including $\kappa\to 0$) and without an azimuthal phase factor, \textit{i.e.}, equivalent to $m_\gamma=\Lambda$ in the preceding equation.

The twisted photon wave front has the feature that the Poynting vector is spiraling forward.  It has azimuthal and $z$ components in cylindrical coordinates, but no radial component.  In detail, the magnetic field for $\Lambda=1$ is 
\begin{align}
B_\rho &=  i\omega \sqrt{\frac{\kappa}{4\pi}} e^{i(k_z z -\omega t + m_\gamma\phi)}
	\nonumber\\
& \quad \times \left( \sin^2\frac{\theta_k}{2} J_{m_\gamma+1}(\kappa\rho) 
		+ \cos^2\frac{\theta_k}{2} J_{m_\gamma-1}(\kappa\rho) \right) \,,	\nonumber\\
B_\phi &=  \omega \sqrt{\frac{\kappa}{4\pi}} e^{i(k_z z -\omega t + m_\gamma\phi)}
	\nonumber\\
& \quad \times \left( \sin^2\frac{\theta_k}{2} J_{m_\gamma+1}(\kappa\rho) 
		- \cos^2\frac{\theta_k}{2} J_{m_\gamma-1}(\kappa\rho) \right) \,,	\nonumber\\
B_z &= \omega \sqrt{\frac{\kappa}{4\pi}} e^{i(k_z z -\omega t + m_\gamma\phi)}
	\sin^2\theta_k J_{m_\gamma}(\kappa\rho)	\,,
\end{align}
and the electric field is just 90$^\circ$ out of phase with the magnetic field, $\vec E = i \vec B$.  The physical fields are the real parts of the above expressions, and one can see the wave front moves forward at less than the normal speed of light.    Working  physical electric and magnetic fields, the Poynting vector $\vec S = \vec E \times \vec B$ is
\begin{align}
S_\rho &= 0	\,,	\nonumber\\
S_\phi &= \frac{\kappa \omega^2}{4\pi} \sin\theta_k \, J_{m_\gamma}(\kappa\rho)
		\nonumber\\
& \quad \times	\left( \cos^2\frac{\theta_k}{2} J_{m_\gamma-1}(\kappa\rho)
		+ \sin^2\frac{\theta_k}{2} J_{m_\gamma+1}(\kappa\rho)	\right)  \,,	\nonumber\\
S_z &= \frac{\kappa \omega^2}{4\pi}
	\left( \cos^4\frac{\theta_k}{2} J^2_{m_\gamma-1}(\kappa\rho)
		- \sin^4\frac{\theta_k}{2} J^2_{m_\gamma+1}(\kappa\rho)	\right)	\,.
\end{align}

\begin{figure}[htbp]
\begin{center}
\includegraphics[width = 78 mm]{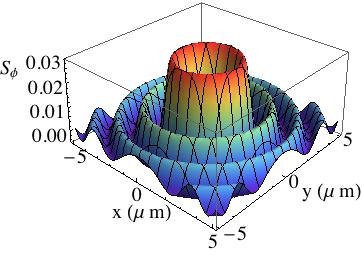}
\caption{The size of Poynting vector azimuthal component as a function of position in the transverse plane.  For this illustration, the photon wavelength is 0.5 microns, the pitch angle is $0.2$ radians, and $m_\gamma=4$. }
\label{fig:sscalar}
\end{center}
\end{figure}

\begin{figure}[htbp]
\begin{center}
\includegraphics[width = 75 mm]{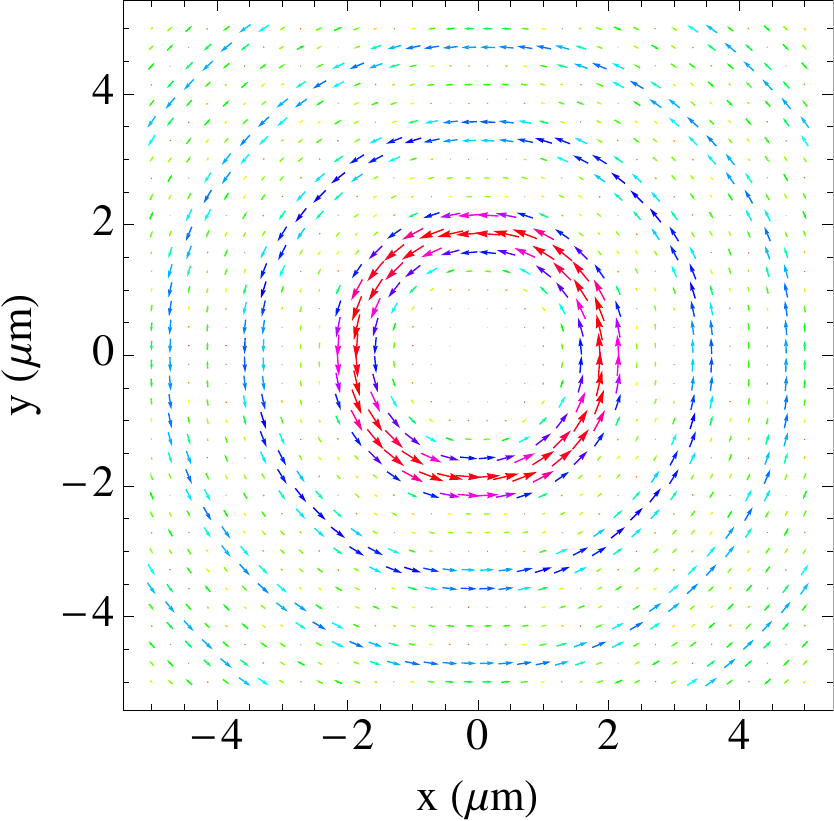}
\caption{A plot of $2\pi\rho$ times $\vec S$ projected onto the transverse plane.  One sees the major bands of $\vec S$ circulating in the same direction, building up the large orbital angular momentum.  Also for this illustration, $\lambda= 0.5 \mu$m, $\theta_k=0.2$ radians, and $m_\gamma=4$.}
\label{fig:svector}
\end{center}
\end{figure}

Figure~\ref{fig:sscalar} shows $S_\phi$ in the transverse plane.  For this illustration, and for the next, the photon wavelength is 0.5 microns, the pitch angle is $0.2$ radians, and $m_\gamma=4$.  The figure shows a bullseye pattern characteristic of twisted photons, with a wide hole in the middle that one can also see from the Bessel functions in the above expressions.  Figure~\ref{fig:svector} shows $\vec S$ in the transverse plane, showing again the bullseye pattern and also showing the circulation of momentum density about the center of the pattern.

The center of the bullseye is currently at the origin in the $x$-$y$ plane.  Shifting it is easily done by applying the translation operator $\exp\{i\hat p{\cdot}b\}$ to the twisted photon state $| \kappa m_\gamma k_z \Lambda \rangle$, where $\hat p^\mu$ is the momentum operator and $b^\mu$ is a constant vector.  There is a \textit{de facto} phase convention in Eq.~(\ref{eq:planewave}), that the momentum eigenstate wave function at $x=0$ is just the polarization vector.  Algebraic effects of the shift are to change the phase in Eq.~(\ref{eq:twistedwave}) from $\exp\{-ikx\}$ to $\exp\{-ik(x-b)\}$ and arguments of the Bessel functions in later equations to $J_\nu(\kappa |\vec x_\perp - \vec b_\perp|)$.


\section{Angular Momentum Projection}			\label{sec:amp}


The factor $\exp\{im_\gamma\phi_k\}$ in the twisted photon state should give the state a $z$-component of angular momentum that is at least approximately $m_\gamma$.  This is stated in a number of sources, but we have not seen a claim of exactness.   In fact we can prove that the total angular momentum projected in the longitudinal direction is exactly $m_\gamma$, at least in the sense of expectation values.

From the Noether current corresponding to rotations, one gets the angular momenta.  The result can be found, for example, in Bjorken and Drell \cite{Bjorken-Drell}, Eq.~(14.22), and is
\be
J^{ij} = \epsilon^{ijk} J^k = \int d^3x  
: \dot{\vec A} \cdot (x^i \partial^j - x^j \partial^i) \vec A - (\dot A^i A^j - \dot A^j A^i) :
\ee
We will speak of the first term as the orbital angular momentum and the second as the spin.  

Regarding the spin term, one can consider a direct calculation with the usual expansion and commutation relation in terms of plane wave states,
\be
A^\mu(x) = \sum_\lambda \int \frac{d^3q}{(2\pi)^3 2\omega_q}
	\left( a_{\vec q \lambda} \epsilon^\mu_{\vec q \lambda} e^{-iqx} + 
	a^\dagger_{\vec q \lambda} \epsilon^{\mu *}_{\vec q \lambda} e^{iqx} \right)	\,,
\ee
and
\be
\big[ a_{\vec q \lambda}, a^\dagger_{\vec k \Lambda} \big] = 
	(2\pi)^3 2\omega \delta^3(\vec q - \vec k) \delta_{\lambda\Lambda}	\,.
\ee

After noting that the $a^\dagger a^\dagger$ and $aa$ terms in $J^3$(spin) do not contribute to the matrix element 
below, one can show the $a^\dagger a$ terms lead to
\begin{align}
\langle k'\Lambda'| J^3({\rm spin}) |k\Lambda \rangle
	= 2i\omega (2\pi)^3 \delta^3(\vec k - \vec k') 
\big( \vec\epsilon_{\vec k \Lambda} \times \vec\epsilon^*_{\vec k \Lambda'} \big)^z  .
\end{align}
After showing
\be
\big( \vec\epsilon_{\vec k \Lambda} \times \vec\epsilon^*_{\vec k \Lambda'} \big)^z
	= -i \Lambda \cos\theta_k \delta_{\Lambda\Lambda'}	\,,
\ee
this becomes
\begin{align}
\langle k'\Lambda'| J^3({\rm spin}) |k\Lambda \rangle
	= \Lambda \cos\theta_k  \ \langle k'\Lambda' |k\Lambda \rangle .
\end{align}
There is no $\phi_k$ dependence above, and since the twisted photon states each have fixed $\Lambda$ and $\theta_k$, one can promote the states to twisted photon states and obtain,
\be
\frac{
\langle \kappa' m_\gamma k'_z \Lambda | J^3({\rm spin})    | \kappa m_\gamma k_z \Lambda \rangle}
{
\langle \kappa' m_\gamma k'_z \Lambda  | \kappa m_\gamma k_z \Lambda \rangle}
	= \Lambda \cos\theta_k 	\,.
\ee

Continuing to the orbital angular momentum (OAM) piece, we need
\begin{align}
&\langle \kappa' m_\gamma k'_z \Lambda | J^3({\rm OAM})  | \kappa m_\gamma k_z \Lambda \rangle
		\nonumber\\
& \quad = \langle \kappa' m_\gamma k'_z \Lambda | 
	\int d^3x : \dot{\vec A} \cdot \frac{\partial \vec A}{\partial \phi_\rho} :
	| \kappa m_\gamma k_z \Lambda \rangle
\end{align}
We pursue a different calculation here, still noting that within the normal ordering terms with two creation or two annihilation operators give zero.  For contributions where an $a_{k \lambda}$ comes from
$\partial \vec A/\partial \phi_\rho$ and an $a^\dagger_{k' \lambda'}$ comes from 
$\dot{\vec A}$, the result is unchanged by inserting a vacuum intermediate state. The same is true for the reverse contribution.  Hence
\begin{align}
&\langle \kappa' m_\gamma k'_z \Lambda | J^3({\rm OAM})  | \kappa m_\gamma k_z \Lambda \rangle
		\nonumber\\
& \quad = 2 \int d^3x  \ 
	\langle \kappa' m_\gamma k'_z \Lambda |  \dot{\vec A} |0 \rangle \cdot 
	\langle 0| \frac{\partial \vec A}{\partial \phi_\rho}
	| \kappa m_\gamma k_z \Lambda \rangle	\,.
\end{align}
We can use the results for the twisted state wave functions, Eq.~(\ref{eq:twistedwf}),  and known Bessel function integrals to obtain
\begin{align}
&\frac{
\langle \kappa' m_\gamma k'_z \Lambda | J^3({\rm OAM})    | \kappa m_\gamma k_z \Lambda \rangle}
{
\langle \kappa' m_\gamma k'_z \Lambda  | \kappa m_\gamma k_z \Lambda \rangle} \nonumber\\
	& \quad = \frac{1}{2} m_\gamma \sin^2\theta_k + (m_\gamma-\Lambda) \cos^4\frac{\theta_k}{2}
	+ (m_\gamma+\Lambda) \sin^4\frac{\theta_k}{2}		\nonumber\\
	&\quad = m_\gamma - \Lambda\cos\theta_k	\,.
\end{align}

Combining the results,
\be
\frac{
\langle \kappa' m_\gamma k'_z \Lambda | J^3    | \kappa m_\gamma k_z \Lambda \rangle}
{
\langle \kappa' m_\gamma k'_z \Lambda  | \kappa m_\gamma k_z \Lambda \rangle}	=	m_\gamma	\,.
\ee
The total angular momentum projection in the direction of motion is precisely $m_\gamma$. The value of $m_\gamma$ can be controlled in the lab by the means the beam of twisted photons is generated, for example, by the use of spiral phase plates or computer-generated holograms \cite{Yao11}.


\section{Atomic photoexcitation}			\label{sec:vector}


We will consider excitation by a twisted photon of a hydrogen-like atom from the ground state to an excited state.

In general, the photon's wave front will be traveling in the $z$-direction and the axis of the twisted photon will be displaced from the nucleus of the atomic target by some distance in the $x$-$y$ plane which we will call $\vec b$.  We work out the photoexcitation for this case in this section, and then shall apply the result to two situations.  

\begin{figure}[b]
\begin{center}
\includegraphics{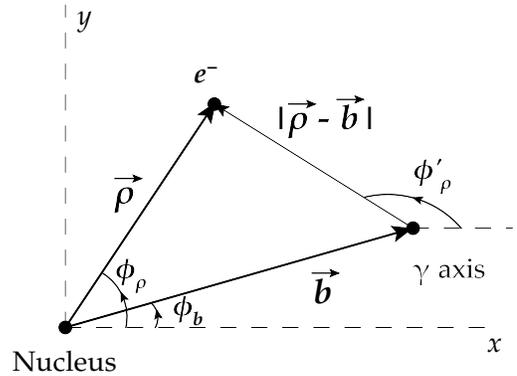}
\caption{Relative positions of atomic state and photon axis, as projected onto the $x$-$y$ plane, with the origin at the nucleus of the atom.}
\label{offaxiscoord}
\end{center}
\end{figure}


One situation will be the case when the twisted photon axis passes directly through the center of the atom's nucleus.  The other will be the case where target atoms are at random locations, and we have to average over all axis to atom separations.

For simplicity, we will treat an atomic state just in terms of its principal quantum number $n_k$, orbital quantum number $l_k$, and magnetic quantum number $m_k$, where $k=i$ for the initial state and $k=f$ for the final state.  

We treat the atom nonrelativistically.  The twisted photon satisfies the Coulomb gauge condition, and the interaction Hamiltonian is
\be
\label{eq:hamiltonian}
H_1 = - \frac{e}{m_e} \vec A \cdot \vec p \,,
\ee
and the transition matrix element is
\begin{align}
S_{fi} &= -i \int dt  \langle n_f l_f m_f | H_1 
	| n_i l_i m_i; \kappa m_\gamma k_z \Lambda \rangle				\,.
\end{align}

The twisted photon wave function is given in Eq.~(\ref{eq:twistedwf}).

\begin{widetext}

We shall center the atomic state at the origin, with the atomic electron located at $(\rho,\phi_\rho,z)$ in cylindrical coordinates or $(r,\theta_r,\phi_\rho)$ in spherical coordinates.  The twisted photon, moving in the $z$-direction, has its origin in general not centered on the atomic nucleus but displaced to position $\vec b$ in the $x$-$y$ plane.  Relative to the photon axis, the electron position projected onto the $x$-$y$ plane will be at distance $| \vec\rho - \vec b |$ and angle $\phi_\rho'$, as illustrated in Fig.~\ref{offaxiscoord}.

Then 
\begin{align}
S_{fi} &= 2\pi i \delta(E_f-E_i-\omega)  \,\frac{e}{m_e}  \sqrt{\frac{\kappa}{2\pi}}	\,
	\int r^2 dr \, d(\cos\theta_r) \, d\phi_\rho \,  
	R_{n_f l_f}(r) Y^*_{l_f m_f}(\theta_r,0) e^{-i m_f \phi_\rho}	
	\Bigg\{
	\frac{\Lambda}{\sqrt{2}} e^{im_\gamma\phi'_\rho} \sin\theta_k \,
	J_{m_\gamma}(\kappa |\vec \rho - \vec b|) \, \vec\eta_0
											\nonumber\\
&\qquad	+  i^{-\Lambda}\,  e^{i(m_\gamma-\Lambda)\phi'_\rho}  \, \cos^2\frac{\theta_k}{2} \,
	J_{m_\gamma-\Lambda}(\kappa |\vec \rho - \vec b|) \, \vec\eta_\Lambda	
+ i^{\Lambda}  e^{i(m_\gamma+\Lambda)\phi'_\rho}  \,  \sin^2\frac{\theta_k}{2}
	J_{m_\gamma+\Lambda}(\kappa |\vec \rho - \vec b|) \, \vec\eta_{-\Lambda}
	\Bigg\}	e^{i k_z z}		\cdot \vec p \, R_{10}(r) Y_{00}		\,,
\end{align}
where $E_k = E_{n_k}$.

Note that,
\begin{align}
\hat\eta_\lambda {\cdot} \vec p \, R_{10}(r) 
	= -i \hat\eta_\lambda {\cdot} \hat r \, R'_{10}(r) =
	-i \sqrt{\frac{4\pi}{3}} Y_{1\lambda}(\theta_r,\phi_\rho) \, R'_{10}(r)
\end{align}

\noindent The expansion theorem,
\begin{align}
e^{i n \phi_\rho'}  J_n(\kappa |\vec \rho - \vec b|)
	= \sum_{N_1=-\infty}^\infty e^{i N_1 \phi_\rho} e^{-i(N_1-n)\phi_b } 
	J_{N_1}(\kappa\rho) J_{N_1-n}(\kappa b)	\,;
\end{align}
allows us to do the $d\phi_\rho$ integral and obtain
\begin{align}
\label{eq:genresult}
S_{fi} &= -2\pi \delta(E_f-E_i-\omega)  \frac{e}{m_e a_0}  \,  
	\sqrt{\frac{2\pi\kappa}{3}}	\,	e^{i(m_\gamma-m_f)\phi_b} \, J_{m_f-m_\gamma}(\kappa b)
					\nonumber\\
&\times	i^{-\Lambda} \Bigg\{
		 \cos^2\frac{\theta_k}{2} \, g_{n_f l_f m_f \Lambda}	+ 		
		 \frac{i}{\sqrt{2}} \sin\theta_k 	\, g_{n_f l_f m_f 0}
	- 	 \sin^2\frac{\theta_k}{2} \,  g_{n_f l_f m_f, -\Lambda}
	\Bigg\}		\nonumber\\
&\stackrel{\rm def}{=} 2\pi \delta(E_f-E_i-\omega)  \, \mathcal M_{n_f l_f m_f \Lambda}(b)
	\,.
\end{align}

\noindent Note that the energy delta-function requires energy conservation, but in our formalism atomic recoil is neglected, so that overall linear momentum is not conserved. The dimensionless atomic factors are
\begin{align}
\label{eq:reduced}
g_{n_f l_f m_f \lambda} \equiv  - a_0
	\int_0^\infty r^2 dr \ R_{n_f l_f}(r)  \, R'_{10}(r)  
	\int_{-1}^1 d(\cos\theta_r) \, J_{m_f-\lambda}(\kappa \rho) \, 
	Y_{l_f m_f}(\theta_r,0) \, Y_{1 \lambda}(\theta_r,0) e^{i k_z z}		\,,
\end{align} 
and $a_0$ is the Bohr radius.  The quantum numbers of the initial state are tacit, as we always start from the ground state.  As a simple practical matter, 
$-a_0 R'_{10}(r) = R_{10}(r)$, and one also has $\kappa\rho = \omega r \sin\theta_r \sin\theta_k$ and $k_z z = \omega r \cos\theta_r \cos\theta_k$.   Further, one can show that the three terms in the curly bracket above are either all real or else all purely imaginary.

\end{widetext}

The magnitude of the result depends on $m_\gamma$ only through the argument of the Bessel function $J_{m_f-m_\gamma}(\kappa b)$.  Hence if the twisted photon axis directly lines up with the atomic nucleus, only the $J_0$ Bessel function is non-zero, and only atomic states whose magnetic quantum number matches the orbital angular momentum of the photon can be excited.  On the other hand, if the twisted photon axis misses the atom's center by a large margin, many Bessel functions are of comparable average magnitude, and the dominantly produced state, if there be one, will be the one with the largest atomic factor $g_{n_f l_f m_f \Lambda}$, with dependence on the photon angular momentum projection $m_\gamma$ being sub-dominant.


\section{On-axis and off-axis atomic excitation}			\label{sec:atomicX}


The axis of the twisted photon may pass directly through the center of the atomic nucleus.   Although perhaps difficult to make happen experimentally, the theoretical result for this case is interesting, particularly because of the applicable selection rules.  

By way of review, the selection rules for photoexcitation (starting from the ground state) with plane wave photons of helicity $\Lambda$ are 
\begin{align}
m_f &= \Lambda,	 \nonumber\\
l_f &\ge 1,		 \nonumber\\
g^{({\rm pw})}_{n_f l_f, m_f = \Lambda, \Lambda} 
	&\propto \left( \omega a_0 \right)^{l_f - 1}  ,
\end{align}
where $g^{({\rm pw})}$ is the plane wave analog of the reduced atomic amplitudes shown in Eq.~(\ref{eq:reduced}), and shows the suppression that follows when higher photon partial waves are needed. Such selection rules are well known from textbooks, cf. Refs. \cite{Schiff,Davydov}.

For a twisted photon striking an atom centered on its axis, the impact parameter $b=0$ and we immediately obtain $m_f = m_\gamma$ from the Bessel function in the general result, Eq.~(\ref{eq:genresult}),
\begin{equation}
J_{m_f-m_\gamma}(\kappa b) \to J_{m_f-m_\gamma}(0)= \delta_{m_f m_\gamma}	\,.
\end{equation}
That is, the only final states that can be produced are those that can absorb the full projected orbital angular momentum of the twisted photon.  The selection rules for the on-axis twisted photoexcitation are,
\begin{align}
m_f &= m_\gamma,		\nonumber\\
l_f &\ge \left| m_\gamma \right|,		\nonumber\\
g_{n_f l_f, m_f = m_\gamma, \Lambda} &\propto \left( \omega a_0 \cos\theta_k \right)^{l_f - 1}
	\left( \tan\theta_k \right)^{ |m_\gamma - \Lambda|}		.
\end{align}
These or analogous selection rules are given for on-axis photoionization in~\cite{Picon10}, respectively. Noteworthy, the authors of Ref. ~\cite{Picon10} used a Laguerre-Gaussian parameterization of the twisted photon states, that is different from a plane-wave expansion \cite{Jentschura:2010ap,Jentschura:2011ih} we used in our calculation; it is re-assuring that we arrived at similar results.
Similar selection rules for the twisted photons aligned with the atomic center were recently derived for the case of elastic forward scattering \cite{Davis13}.

\begin{figure}[htbp]
\begin{center}
\includegraphics[width = 84 mm]{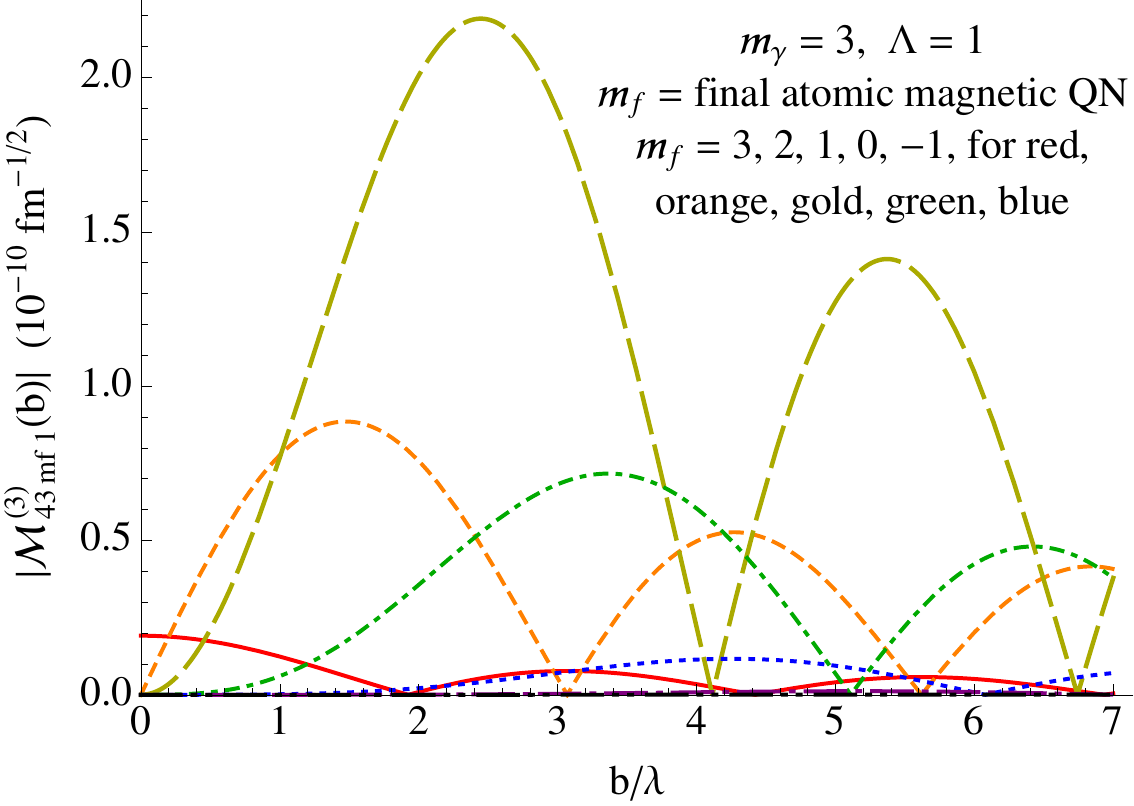}
\includegraphics[width = 84 mm]{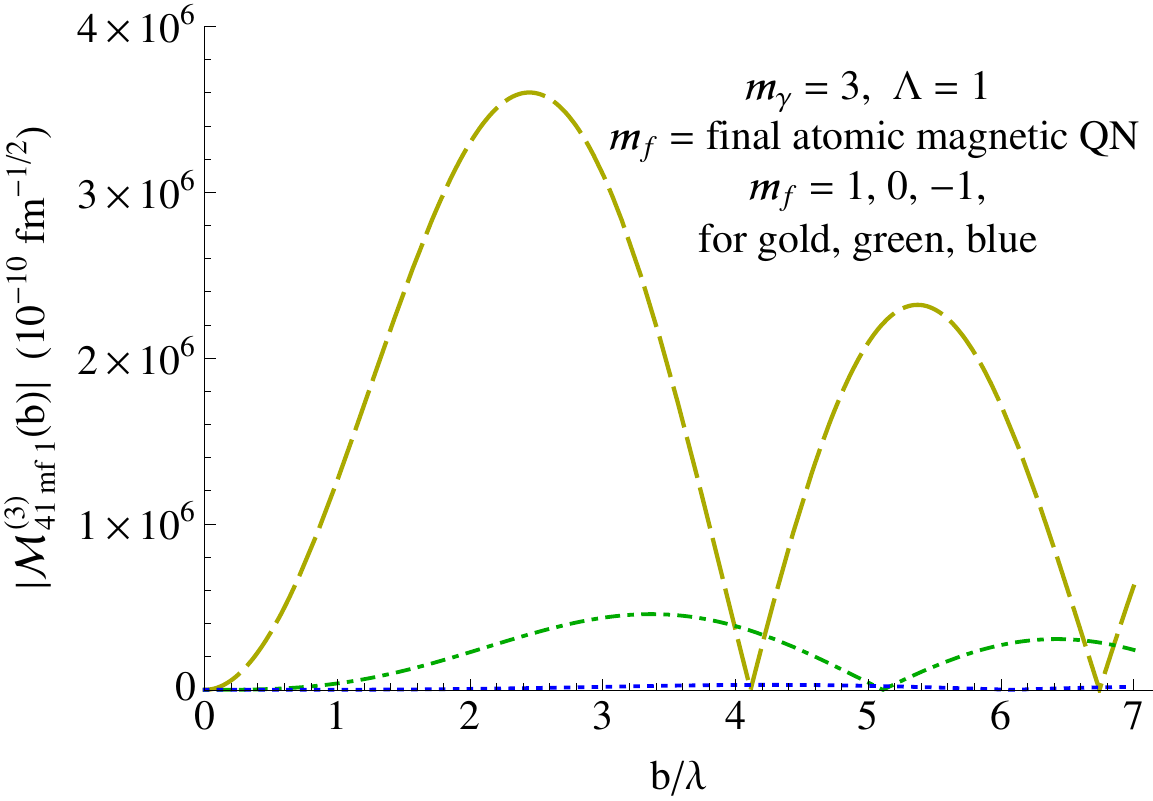}
\caption{Size of the transition amplitude $\left|  \mathcal M^{(m_\gamma=3)}_{n_f l_f m_f \Lambda}(b)  \right|$ for particular quantum numbers $n_f, l_f,\Lambda$ and several $m_f$.  On the upper graph, $m_f=3$ is the red solid curve, $m_f=2$ is orange and medium dashed, $m_f=1$ is gold and long dashed, $m_f=0$ is green and dot-dashed, $m_f=-1$ is blue and dotted, and transitions to other $m_f$ are quite small and not plotted. Lower graph is for the final state $l_f=1$; the state $m_f=1$ is allowed by electric-dipole selection rules for plane waves, while $m_f=0,-1$ are unique for the twisted photons.}
\label{fig:impact}
\end{center}
\end{figure}

However, the atom does not have to be far off the photon axis before other amplitudes, not satisfying the above selection rules, play an important role.  As illustration, we plot in Fig.~\ref{fig:impact} the amplitudes $\left|  \mathcal M_{n_f l_f m_f \Lambda}(b)  \right|$ for the example of $n_f=4$, $l_f = 3$, $\Lambda=1$, photon energy and wavelength set by the H-atom level spacing, and photon angular momentum along the direction of motion $m_\gamma=3$ (upper plot) and $n_f=4$, $l_f = 1$, $\Lambda=1$ (lower plot). Note the relative strength of the amplitudes is much higher, by about six orders of magnitudes, for the transition into $l_f=1$ state vs $l_f=3$, in accordance with the selection rules presented here.

The horizontal axis is the impact parameter in units of the photon wavelength.  Already with an impact parameter of less than half a wavelength, amplitudes that do not satisfy the $m_f = m_\gamma$ selection rule are becoming important.  The amplitude with the largest peak is the one with $m_f=1$, which is the only amplitude one would have with a plane-wave photon polarized with helicity $\Lambda = 1$.  

The general selection rules for the off-axis case are
\begin{align}
m_f &= {\rm any},		\nonumber\\
l_f &\ge \left| m_f \right|	,	\nonumber\\
g_{n_f l_f, m_f = m_\gamma, \Lambda} &\propto \left( \omega a_0 \cos\theta_k \right)^{l_f - 1}
	\left( \tan\theta_k \right)^{ |m_f - \Lambda|}	.
\end{align}

Above selection rules may seem to be in conflict with conservation of total angular momentum. Note, however that since the absorbed photons have a momentum not aligned with z-axis and we neglect atomic recoil, momentum is no longer conserved and rotational symmetry with respect to z-axis does not hold anymore, as opposed to an on-axis case. Conservation of total angular momentum can be restored if the recoil momentum of an atom is taken into account.


\section{Cross sections for randomly distributed targets}	\label{sec:xsctn}



\subsection{Cross section calculation}			\label{sec:xsc}


 In general, if a photon passes through a target containing many atoms, one cannot expect that the location of the photon axis can be controlled at the level of the atomic spacing.  Hence, we should average over the transverse separations, when calculating the photoexcitation cross sections.

\begin{figure}[tbp]
\begin{center}
\includegraphics{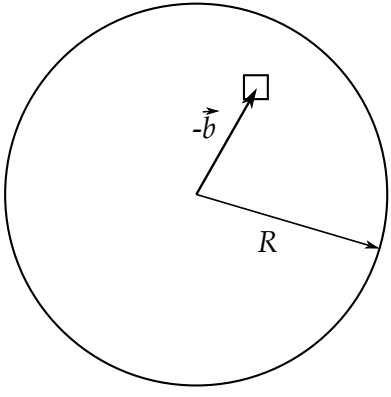}
\caption{A large target with circular cross section of radius $R$, centered on the twisted photon axis, with a target atom at location $-\vec b$.}
\label{circulartarget}
\end{center}
\end{figure}

It will be convenient to think of a twisted photon incident on the center of a large target of circular cross section and radius $R$, with a particular target atom at a random transverse location $-\vec b$ (to match our previous considerations), as in Fig.~\ref{circulartarget}.


For a given $\vec b$, 
\begin{equation}
\sigma_{n_f l_f m_f \Lambda} = 2 \pi \delta(E_f -E_i-\omega_\gamma)
 \frac{ |\mathcal M_{n_f l_f m_f \Lambda}(b)|^2 }{f} \,.
\end{equation}
where $f$ is the incoming flux and a suitable sum or average over spins is implied.  

With the target atoms uniformly distributed, the cross section averaged over atom location is
\begin{align}
&\overline\sigma_{n_f l_f m_f \Lambda} 
	=  \frac{ 2 \pi \delta(E_f -E_i-\omega) }{f}
					\nonumber\\
&\hskip 3.8 em	\times   \frac{1}{\pi R^2} \int d^2b \ 
		|\mathcal M_{n_f l_f m_f \Lambda}(b)|^2	
					\nonumber\\[1.2ex]
&= \frac{ 2 \pi \delta(E_f -E_i-\omega) }{f}
	\frac{1}{\pi R^2}	 \frac{ 2\pi e^2 \kappa }{3 m_e^2 a_0^2}
											\nonumber\\
&\,\times	\int_0^R 2\pi b \, db \  	J^2_{m_f-m_\gamma}(\kappa b)
\times	
	\Big|
	\cos^2 \frac{\theta_k}{2} \,	g_{n_f l_f m_f \Lambda}
											\nonumber\\
&\quad	+ \frac{i}{\sqrt{2}} \sin\theta_k 	\, g_{n_f l_f m_f 0}
	-	\sin^2\frac{\theta_k}{2} \,	g_{n_f l_f m_f, -\Lambda}  \Big|^2
\,.
\end{align}

\noindent Outside the integration measure, the only $b$ dependence is in the Bessel function, and the useful integral is
\begin{equation}
\lim_{R\to\infty} \int_0^R  b \, db \ J^2_{m_f-m_\gamma}(\kappa b) 
	= \frac{R}{\pi \kappa}		\,,
\end{equation}
independent of index.

For the flux we take the average density of the twisted photon state times the incoming wave front velocity $k_z/\omega$.  The target is unit normalized.  The density of the twisted photon state, with our normalization and averaged over a disk of radius $R$, can be worked out and leads to
\begin{equation}
f = \rho_{\rm avg}\frac{k_z}{\omega} = \frac{2 k_z}{\pi^2 R}		\,.
\end{equation}

Thus
\begin{align}
\overline\sigma_{n_f l_f m_f \Lambda} &= 2\pi \delta(E_f -E_i-\omega)
	\frac{ 8\pi^3 \alpha^3  }{3  k_z} \  \bigg|
	\cos^2 \frac{\theta_k}{2} \,	g_{n_f l_f m_f \Lambda}
											\nonumber\\
&\quad	+ \frac{i}{\sqrt{2}} \sin\theta_k 	\, g_{n_f l_f m_f 0}
	-	\sin^2\frac{\theta_k}{2} \,	g_{n_f l_f m_f, -\Lambda}  \bigg|^2
\,.		
\end{align}


\subsection{Unique twisted photon features}			\label{sec:tpf}


For the perfectly centered on target twisted photon, there is the dramatic result that the magnetic quantum number of the final atomic state must equal the corresponding z-projection of orbital angular momentum of the twisted photon.  

For the general case of random target location, there are still features unique to twisted photons.  States can be photoexcited that cannot be reached by plane-wave photons.  However, as we will demonstrate below in the numerical examples, the overall cross sections will not be much changed compared to plane-wave photons.

Photoexcitation, starting from the ground state, by a plane-wave photon of a certain helicity leads only to final states whose magnetic quantum number equals that helicity.  One can work out photoexcitation of a state $(n_f,l_f,m_f)$ by a plane-wave photon of helicity $\Lambda$ starting from the same Hamiltonian, Eq.~(\ref{eq:hamiltonian}), and express the result as
\begin{align}
\sigma_{n_f l_f m_f \Lambda}^{(pw)} = 2\pi \delta(E_f-E_i -\omega)	
	\frac{8\pi^3\alpha^3}{3\omega}  \delta_{m_f \Lambda}	\,
	g^{(0)}_{n_f l_f \Lambda \Lambda}		\ ,
\end{align}
where
\begin{align}
g^{(pw)}_{n_f l_f \Lambda \Lambda} &= - a_0 
	\int_0^\infty r^2 dr \, R_{n_fl_f}(r) R'_{10}(r)
				\nonumber\\
&\times \int_{-1}^1 d(\cos\theta_r)	\,
	 Y_{l_f \Lambda}(\theta_r,0) Y_{1\Lambda}(\theta_r,0) e^{i\omega z}   \,,
\end{align}
which is identical to the $\theta_k \to 0$ limit of the twisted photon result.

Twisted photons, on the other hand, photoexcite states with a large range of magnetic quantum numbers $m_f$.  Values of $m_f$ impossible for plane-wave photons are produced even when the twisted photons enter a medium with random target locations.

Twisted photons also produce the $m_f=\Lambda$ states that plane-wave photons necessarily lead to.  But the interest is in the $m_f \ne \Lambda$ states unique to twisted photon production.  To quantify the probability of finding these states, we define a ratio which compares the rate for producing final states that are unique to twisted photons to the total rate where the twisted photon produces all final states, including $m_f = \Lambda$, for a given energy level characterized by quantum numbers $(n_f,l_f)$ (and for the case of a large interaction region with random target locations),
\begin{align}
f_{\rm twisted} = 
	\frac{\sum_{\stackrel{\scriptstyle{m_f = - l_f,}}{m_f\ne \Lambda}}^{m_f=l_f}
	\overline\sigma_{n_f l_f m_f \Lambda}}
	{\sum_{m_f = - l_f}^{m_f=l_f}
	\overline\sigma_{n_f l_f m_f \Lambda}}	  \,.
\end{align}
Here we have fixed $\Lambda$.  We could also average over $\Lambda$ in the case of unpolarized photons.

Another ratio of potential interest is the comparison between the total photoproduction rate from twisted photons and the corresponding result for plane-wave photons,
\begin{equation}
r_{\rm twisted} = \frac{ \sum_{m_f = - l_f}^{m_f=l_f}
	\overline\sigma_{n_f l_f m_f \Lambda} } 
	{\sigma_{n_f l_f \Lambda \Lambda}^{(pw)}}
\end{equation}

One may also compare the probabilities of photoexcitation of an (unpolarized) atom by photons with opposite helicities $\pm\Lambda$. For plane-wave photons these probabilities are identical due to parity conservation. For twisted photons with a fixed $z$-projection of orbital angular momentum but opposite helicities such an asymmetry would not violate parity, because the corresponding photon states do not transform into each other via parity transformation. Calculations show that such an asymmetry is indeed nonzero for a fixed value of an impact parameter $b$; however, it turns to zero once we average over $b$.


\subsection{Numerical examples}			\label{sec:numex}


The "twisted photon ratio," $f_{\rm twisted}$, evaluates what fraction of the final states excited by the twisted photon could not have been produced by a plane-wave photon.  As a numerical example, we evaluate this ratio for final states with varied values of $n_f$ and  $l_f$ and for a twisted photon with pitch angle of $\theta_k=0.2$ radians.  The energy is fixed by the H-atom level spacing, corresponding to the photon wavelength of 100nm.   

First we consider a transition into $n_f=4$, $l_f=1$ state. This is the largest transition, and standard electric-dipole selection rules for plane waves permit transition into
$m_f=1$ (for our choice of photon helicity $\Lambda=1$). For the twisted photons, transitions into $m_f=0,-1$ are also allowed. Taking $m_\gamma=3$, we find for the above choice of parameters:

\begin{equation}
f_{\rm twisted}\left[ {\rm gnd.\ state} \to (4, 1) \right] = 2.0\%,	\,,
\end{equation}
\noindent which indicated the fraction of transitions not accessible for plane-wave photons.

It is instructive to compare the result is for higher orbital angular momentum of atomic states. For $n_f=4, l_f=3$ we get
\begin{equation}
f_{\rm twisted}\left[ {\rm gnd.\ state} \to (4, 3) \right] = 20.3\%.	\,,
\end{equation}

\noindent A similar evaluation with $n_f=5, l_f=4$ yields

\begin{equation}
f_{\rm twisted}\left[ {\rm gnd.\ state} \to (5, 4) \right] = 33.0\%	\,.
\end{equation}

One can see that the fraction of transitions that are unique to the twisted photon is increasing with $l_f$. 
Overall probability of transitions to the states with higher $l_f$ is suppressed according to the selection rules of Eq. (34).

The relative production rate for twisted photons compared to plane-wave ones is measured by the ratio $r_{\rm twisted}$. For all the final states considered above this ratio worked out to $r_{\rm twisted} = 1.02$.  There is a modest enhancement in the photon interaction rate from using twisted photons.  One may check that this is mainly due to an $\omega/k_z$ factor following from the difference in flux factors, which in turn follows from the slight slowing of the wave front's forward motion for twisted photons.


\section{Summary and Discussion}			\label{sec:disc}


In this paper we developed a formalism for photoexcitation of an atom with a beam of twisted photons, using a hydrogen atom as an example. In the derivation, we use an expansion \cite{Jentschura:2010ap,Jentschura:2011ih} of the twisted photon states in terms of plane waves. We show that in a special case when the photon beam axis coincides with the atomic center, the transitions between atomic levels obey angular momentum selection rules, similar to the conclusions made in Ref.\cite{Picon10} for photoionization, resulting in an excited state with a magnetic quantum number ($J_z=m_f$) exactly matching $m_\gamma$ of the incoming beam. We also recover standard angular momentum selection rules for electric-type photon absorption \cite{Schiff, Davydov} in the limit of plane-wave photons. We note, however, that the probability to excite the states with higher orbital angular momentum $l_f$ is suppressed for twisted photons by a familiar factor $(\omega a_0)^{2 (l_f-1)}$ known from the selection rules for the plane-wave photons. This is a significant suppression that corresponds to about six orders of magnitude for the change in $l_f$ by one unit (for visible light); such a suppression is due to non-relativistic nature of electron motion in atoms.

Next, we extended our calculation to a more general case of the atoms located away from the photon beam axis and analyzed the amplitudes of various transitions as a function of the beam center position in units of photon wavelength, $b/\lambda$. In this case the magnetic quantum number of the photoexcited state no longer matches $m_\gamma$, and a range of final-state quantum numbers $m_f$ is generated. Relative magnitudes of various transitions were studied both analytically and numerically. It was found that after we average over the beam position, the amplitude allowed by standard plane-wave selection rules quickly gains strength and makes a dominant contribution to photoabsorption. Still, probabilities of the transitions to higher-$l_f$ states remain suppressed by the factor of $(\omega a_0)^{2 (l_f-1)}$. 

Our calculations show that twisted photons that carry large angular momentum are not more efficient in excitation of higher-$l_f$ atomic states than usual plane-wave photons.
However, the twisted photons produce a range of magnetic quantum numbers $|m_f|\leq l_f$, while for a plane-wave photon of the same helicity +1 only $m_f=1$ is allowed by selection rules.  We introduced several observables that describe probabilities of excitation of the $m_f\neq \Lambda$ states that are forbidden to plane waves. The important finding is that relative probability of transitions to "forbidden" states can reach  a few per cent.

Given a noticeable effect arising from the unique features of the the twisted photons, our predictions can be verified experimentally. 









\begin{acknowledgments}

CEC thanks the National Science Foundation for support under Grants PHY-0855618 and PHY-1205905. Work of AA was supported by The George Washington University. AM thanks JLab and College of William and Mary for hospitality and support during initial stages of this work.

\end{acknowledgments}


\bibliography{TwistedPhoton}

\end{document}